\documentclass[10pt]{article}
\usepackage{natbib}
\usepackage{authblk}
\usepackage[textfont = {footnotesize} , labelfont = bf]{caption}
\usepackage{subfig}
\usepackage{graphicx}
\usepackage{setspace}
\usepackage{url}  
\usepackage{ifthen}  
\usepackage{multicol}   
\usepackage{amsmath}
\usepackage{amsfonts}
\usepackage{natbib}
\usepackage{subfig}
\usepackage{graphicx}
\usepackage{epstopdf}
\usepackage{algorithm}
\usepackage{algorithmic}
\usepackage[margin=1in]{geometry}
 \def \iPAC{\emph{iPAC}}
\def \GraphPAC{\emph{GraphPAC}}
\def \NMC{\emph{NMC}}
\def \SpacePAC{\emph{SpacePAC}}
\def \CHASM{\emph{CHASM}}

\begin{document}

\title{A Spatial Simulation Approach to Account for Protein Structure When Identifying Non-Random Somatic Mutations}
 
\author[1]{Gregory Ryslik}
\author[2] {Yuwei Cheng}
\author[2,3]{Kei-Hoi Cheung}
\author[4]{Robert Bjornson}
\author[1]{Daniel Zelterman}
\author[5]{Yorgo Modis}
\author[1,2]{Hongyu Zhao}

\affil[1]{Department of Biostatistics, Yale School of Public Health, New Haven, CT, USA}
\affil[2]{Program of Computational Biology and Bioinformatics, Yale University, New Haven, CT, USA}
\affil[3]{Yale Center for Medical Informatics, Yale School of Medicine, New Haven, CT, USA}
\affil[4]{Department of Computer Science, Yale University, New Haven, CT, USA}
\affil[5]{Department of Molecular Biophysics \& Biochemistry, Yale University, New Haven, CT,USA}

\maketitle

\begin{abstract}
\noindent \textbf{Background:}
Current research suggests that a small set of ``driver" mutations are   responsible for tumorigenesis while a larger body of ``passenger" mutations occurs in the tumor but does not progress the disease. Due to recent pharmacological successes in treating cancers caused by driver mutations, a variety of of methodologies that attempt to identify such mutations have been developed. Based on the hypothesis that driver mutations tend to cluster in key regions of the protein,  the development of cluster identification algorithms has become critical. 

\noindent \textbf{Results:} 
We have developed a novel methodology, \SpacePAC{} (\textbf{S}patial \textbf{P}rotein \textbf{A}mino acid \textbf{C}lustering), that identifies mutational clustering by considering the protein tertiary structure directly in 3D space. By combining the mutational data in the Catalogue of Somatic Mutations in Cancer (COSMIC) and the spatial information in the Protein Data Bank (PDB), \SpacePAC{} is able to identify novel mutation clusters in many proteins such as FGFR3 and CHRM2. In addition,  \SpacePAC{} is better able to localize the most significant mutational hotspots as demonstrated in the cases of BRAF and ALK. The R package is available on Bioconductor at: \url{http://www.bioconductor.org/packages/release/bioc/html/SpacePAC.html}.

\noindent \textbf{Conclusion:} \SpacePAC{} adds a valuable tool to the identification of mutational clusters while considering protein tertiary structure.

\end{abstract}

\section{Background}
Cancer, at its most basic, is caused by the accrual of somatic mutations within oncogenes and tumor suppressors in the genome \citep{vogelstein_cancer_2004}. While mutations within tumor suppressors usually lower or completely disrupt the activity of genes that promote cell apoptosis or regulate the cell cycle, oncogenic mutations typically increase or destabilize the resulting protein output. As it is easier to disrupt protein function than restore it, there has been significant pharmacological research geared towards inhibiting oncogenic mutations as described in \citet{faivre_current_2006, hartmann_tyrosine_2009} and \citet{moreau_proteasome_2012}. Coupled with the idea of ``oncogene addiction", that a small set of ``driver" genes promote uncontrolled cellular growth in a wide variety of cancers and that inactivation of these genes can significantly impair tumorigenesis \citep{greenman_patterns_2007, weinstein_mechanisms_2006}, the identification of driver oncogenic mutations has become of key importance due to its potential translational benefit. 

Due to the biological importance of this problem, a variety of methodologies have been proposed to identify regions where activating mutations may occur. One approach is based on the idea that compared to the background mutation rate, driver mutations will have a higher frequency of non-synonymous mutations \citep{sjoblom_consensus_2006, bardelli_mutational_2003}. Several improvements to this approach have been made such as normalizing for gene length \citep{wang_prevalence_2002} as well as accounting for different mutation rates due to features such as transitions versus transversions, location of \emph{CpG} sites and tumor type \citep{youn_identifying_2010}. Relatedly, instead of comparing the mutational frequency directly to the background rate, one can also compare the ratio of nonsynonymous ($K_a$) to synonymous ($K_s$) mutations per site \citep{kreitman_methods_2000}. Recently, \citet{bardelli_mutational_2003, lynch_activating_2004, greenman_patterns_2007} and \citet{torkamani_prediction_2008}  showed that somatic mutations appear to cluster within protein kinases while  \citet{wagner_rapid_2007} and \cite{zhou_detecting_2008} demonstrated that mutational clustering can be a sign of positive selection for protein function and thus sites for protein engineering.  

Alternatively, several machine learning methods have been developed to determine the nature of a specific mutation. For instance, \emph{Polyphen-2} \citep{adzhubei_method_2010} attempts to discern whether a mutation is deleterious while \emph{CHASM} \citep{carter_cancer-specific_2009} attempts to distinguish between driver and passenger mutations. These methods use a wide range of sequence and non-sequence features to build a set of rules that are then used to score each mutation with the score value determining the mutation classification. The rules are developed on a training data set via a variety of methods such as Random Forests \citep{breiman_randomforest}, Bayesian Networks \citep{friedman1997bayesian} and Support Vector Machines \citep{cortes_support-vector_1995}.  The features used often include information on the size and polarity of the substituted and original residues, available structural information as well as evolutionary conservation \citep{reva_predicting_2011}. Some classifiers  are optimized to use only a small set of features  in prediction. For example,  the \emph{SIFT} classifier \citep{ng_predicting_2001}, only uses evolutionary conservation to predict whether the protein functional change is tolerated or damaging.


While all these methods have shown some success in identifying damaging or deleterious mutations, they nevertheless have limitations. Methods that rely upon differentiating the frequency of synonymous and non-synonymous mutations as compared to the background rate may fail to take into account that selection may occur upon only a small region of the gene and that signal loss may occur when the gene is considered in total. The methodologies proposed by \citet{wang_prevalence_2002} and \citet{kreitman_methods_2000} fail to distinguish between activating and non-activating non-synonymous mutations while the method developed by \citet{youn_identifying_2010} may be biased if the background mutational rate is not accurately estimated. Furthermore, not only do machine learning classifiers require several sources of information that need to be periodically updated to account for new research, it is often the case that much of the requisite information needs to be collected for the first time at significant expense.

Building upon the hypothesis that driver mutations tend to cluster in functionally relevant protein regions, \citet{ye_2010, ryslik_2013} and \citet{GraphPAC_2013} recently developed several statistical methodologies to identify mutational clusters.  Specifically, \citet{ye_2010} developed Non-Random Mutational Clustering (\NMC{}) which identifies mutational clusters by testing against the null hypothesis that missense substitutions follow a uniform distribution. 
\iPAC{} \citep{ryslik_2013} and \GraphPAC{} \citep{GraphPAC_2013} expanded upon \NMC{} by taking into account protein tertiary structure via a MultiDimensional Scaling approach (MDS) \citep{borg_modern_1997} and a graph theoretical approach, respectively. While both of these methods improved over the linear \NMC{} method, they nevertheless remap the protein to one dimensional space resulting in information loss.

In this article, we provide an improvement to \iPAC{}, \GraphPAC{} and \NMC{} by considering the protein directly in three dimensional space and thereby avoid the information loss inherent in dimension reduction algorithms. Using this new approach, we identify the most number of proteins with significant clusters at the same false discovery rate, including proteins, such as FGFR3 and CHRM2,  whose clusters are missed by \iPAC{}, \GraphPAC{} and \NMC{} (see Section \ref{new}). In addition, \SpacePAC{} provides better ``localization" for mutational hotspots (see Section \ref{localization}). Finally, we show that many of the mutational hotspots identified by \SpacePAC{} are categorized as  activating mutations by \emph{CHASM} and damaging mutations by \emph{PolyPhen-2}. Overall, by avoiding the protein remapping step required by \iPAC{} and \GraphPAC{} as well as the multiple comparison penalty these methods incur for looking at every pairwise combination of mutations, we are better able to identify mutational hotspots that are indicative of driver mutations. For the rest of this paper, we refer to the set of \NMC{}, \GraphPAC{}, and \iPAC{} as the ``pairwise methods" as they consider every pairwise combination at the cost of an extra multiple comparison adjustment.

\section{Methods}
\SpacePAC{} uses a three step process to identify mutational clusters. Step one is to obtain the mutational and structural data (see Sections \ref{data:COSMIC} and \ref{data:PDB}). Step two is to reconcile the databases so that the mutational information can be mapped onto the protein structure (see Section \ref{data:Reconciliation}). The third step is to simulate the distribution of mutation locations over the protein tertiary structure and identify if any regions of the protein have observed mutational counts in the tail of the distribution (see Section \ref{Simulation}). Finally, although not part of the \SpacePAC{} algorithm, we perform a multiple comparison adjustment to account for the multiple structures considered (see Section \ref{MultCompStruct}).

\subsection{Obtaining Mutational Data} \label{data:COSMIC}

The 65th version of the COSMIC SQL database, the latest version as of when this paper was written, was used to obtain the mutation data.  First, only missense substitution mutations recorded as ``confirmed somatic variant" or ``Reported in another cancer sample as somatic" were used. Further as both \SpacePAC{} and the pairwise methods  are tested against the null hypothesis that mutations are randomly distributed along the protein, only mutations from whole gene or whole genome screens were retained in order to avoid selection bias.  Next, only genes labeled with a Uniprot Accession Number \citep{the_uniprot_consortium_reorganizing_2011} were kept as the Uniprot identification was used to match the protein sequence to the protein structure. Finally, as several studies may report mutational data from a single cell line, mutation duplications were removed in order to avoid over counting specific mutations (see ``COSMIC Query" in the supplementary materials for the SQL code).

\subsection{Obtaining 3D Structural Data} \label{data:PDB}
Protein structures were obtained from the PDB. The spatial coordinates of the $\alpha$-carbon atom in each amino acid were used to represent that amino acid's location in 3D space. Further, as multiple structures are often available for the same protein, all structures that matched the protein's UniProt Accession Number were analyzed and an appropriate multiple comparison adjustment applied afterwards (see Section \ref{MultCompStruct}). If multiple polypeptide chains within the same structure matched the Uniprot Accession Number, the first matching chain shown in the file was used (commonly chain ``A"). Similarly, if the structure resolution provided more than one protein conformation, the first conformation listed in the pdb file was kept. For a full listing of the 1,903 structure/side-chain combinations considered, see ``Structure Files" in the supplementary materials.

\subsection{Reconciling Structural and Mutational Data} \label{data:Reconciliation}

As the residue numbering scheme often differs between the COSMIC and PDB databases, we reconciled the information in both sources. Similar to \iPAC{} and \GraphPAC{}, \SpacePAC{} provides the user two possible reconciliation options. The first option is based upon a numerical reconstruction from the structural data available directly in the PDB file while the second performs a pairwise alignment as detailed in \citet{Biostrings_2012}. As the PDB file structure may change depending upon when the file was added to the database along with other technical difficulties, we used pairwise alignment to reconcile the mutational and positional information for each structure unless specifically noted. For further information on the pairwise alignmnent, please see the \iPAC{} package available on Bioconductor. Successful alignment was obtained for 131 proteins corresponding to 1,110 individual structure/side-chain combinations. Note that structures that did not have tertiary data on at least two mutations were considered blank (as no clustering is possible) and dropped from the analysis. See ``Structure Files" in the supplementary materials for full details of each combination.

\subsection{Identifying Mutational Hotspots} \label{Simulation}

The general principle for \SpacePAC{} is that we identify the one, two and three non-overlapping spheres that cover as many of the mutations as possible at various radii lengths. We then normalize the number of mutations covered by the spheres and find the maximum normalized value. This value is then compared to a simulated distribution to obtain a p-value. Specifically, we proceed as follows:
\begin{itemize}
\item Let s be the number of spheres we consider. $s \in \{1,2,3\}$.
\item Let r be the radius considered. Here we consider, $r \in \{1,2,3,4,5,6,7,8,9,10\}$.
\item Simulate $T (\geq 1000)$ distributions of mutation locations over the protein structure. For each simulation, the mutations are randomly permuted among all the amino acids.
\end{itemize}

\begin{figure}[htb!]
\begin{center}
\includegraphics [width = 120mm] {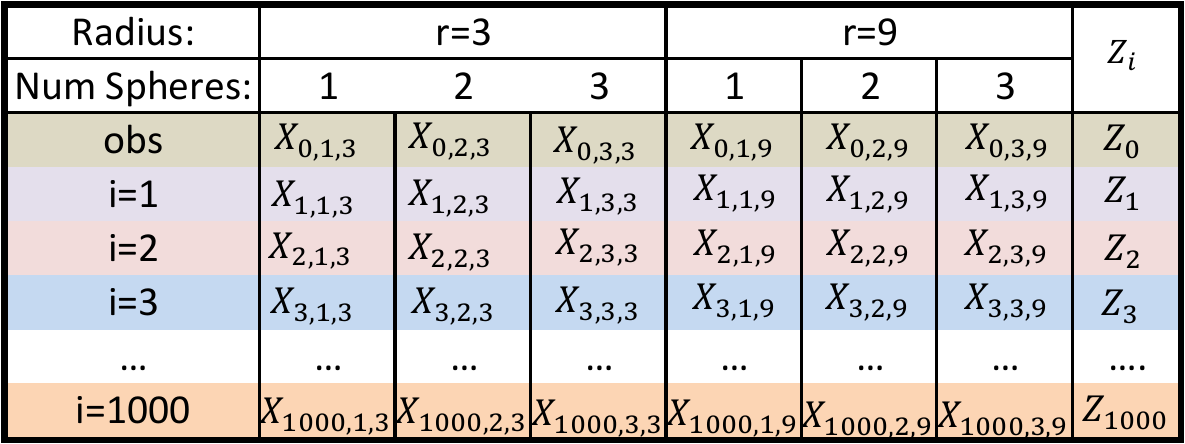}
\caption{In this example, we consider $r \in \{3,9\}$ and up to 3 potential mutational hotspots in the protein. First, $\mu$ and $\sigma$ are calculated over each column. Next, we normalize each entry in the column by calculating $Z_{i,s,r} = \frac{ X_{i,s,r }-  \mu_{s,r}}{\sigma_{s,r}}$. We then take the maximum over each row to get $Z_0,..., Z_{1000}$. The percentage of times $Z_0 \geq Z_i$, where $i \in \{1,...,1000\}$, is the p-value of our observed statistic $Z_0$. Note that if $Z_0$ is less than $Z_i$ for all 1000 simulations, we report a p-value of $<$1.00 E-03.}
\label{FigWork}
\end{center}
\end{figure}

Next, let $X_{0,s,r}$ represent the number of mutations captured by the $s$ spheres. If $s=3$, we identify the sphere centers, $p_1, p_2, p_3$, in such a way such that the spheres capture as many of the total mutations as possible (See Section \ref{quickly}). Let $X_{i,s,r}$ represent the same statistic but for simulation $i$. For given $\{s,r\}$, calculate $\mu_{s,r} = \underset{1 \leq i \leq T}{\operatorname{mean}}\{X_{i,s,r}\}$ and $\sigma_{s,r} = \underset{1 \leq i \leq T}{\operatorname{std.dev.}}\{X_{i,s,r}\}$. For each simulation, calculate  $Z_i = \max_i \{ ( X_{i,s,r}- \mu_{s,r})/\sigma_{s,r} \}$. The p-value is then estimated as:  $(\sum{\mathbf{1}_{Z_{0} > Z_{i}}}) / T$. Note that while  we identify up to three spheres (or ``hotspots") that contain mutational clustering, only one p-value is necessary to reflect the statistical significance of all the hotspots. This process is best seen through Figure \ref{FigWork}. For the rest of this paper, we will refer to ``hotspot" and ``cluster" interchangeably.

\subsection{Algorithm for identifying sphere positions} \label{quickly}
In the approach described in Section \ref{Simulation}, we find the 1, 2 and 3 non-overlapping spheres that cover the most mutations given a pre-specified radius length. Ignoring sphere overlap for the moment, if only one sphere is considered, the number of  possible spheres is linear in the length of the protein (namely, a sphere centered at each residue). If two spheres are considered, there are $N \choose 2$ possible sphere combinations if the protein is $N$ residues long. Similarly, if three spheres are considered, there are $N \choose 3$ such combinations (once again ignoring sphere overlap). For a medium-sized protein like PIK3C$\alpha$ which is 1,068 residues long, considering three spheres allows for $1,068 \choose 3$  $= 202,461,116$ possible positions. This makes it prohibitively expensive to perform a brute force approach. To quickly find the best sphere orientation, we execute Algorithm \ref{codesnip} below. Algorithm \ref{codesnip} is presented for 2 spheres but is trivially extendable to 3 or more spheres as well. 

{\fontsize{5}{5}\selectfont
\begin{algorithm}
\caption{We are interested in finding the two non-overlapping spheres that contain the most mutations. The algorithm takes as input a sorted vector $v$ of mutation counts per amino acid where the amino acids are sorted from largest to smallest by mutation count. Amino acids with no mutations are not included in $v$. Note that the ``cand" variable is an ordered list of 3-tuples.}
\label{codesnip}
\begin{algorithmic}                    
    \REQUIRE Sorted vector of counts $v$ with length $>=2$
    \STATE starti = 2;
    \STATE startj = 1;
    \STATE k = length(v);
    \STATE cand = [(starti, startj,v[starti] + v[startj])]
    \WHILE{ (!is.empty(cand))}
    \STATE index = max(cand) \COMMENT{\#Max over the 3rd element in the 3-tuple.}
	\STATE i,j,s = cand[index]
    \STATE cand = cand[-index] \COMMENT{\#Removes the current max.}

		\IF { (No overlap between sphere i and j)}
			\STATE Return (i, j, v[i]+v[j]) \COMMENT{\#Successful combination found.}
		\ENDIF		
		
		\IF {(j ==1) and ($i < k$)}
			\STATE cand.append[(i+1, j,v[i+1] + v[j])]
		\ENDIF
		\IF {($j +1 < i$)}
			\STATE	cand.append[(i, j+1,v[i] + v[j+1])]
		\ENDIF

    \ENDWHILE
  
  	\STATE Return NULL \COMMENT{\#No succesful combination found.}
\end{algorithmic}
\end{algorithm}
}

\begin{figure}[h!]
\begin{center}
\includegraphics [width = 120mm] {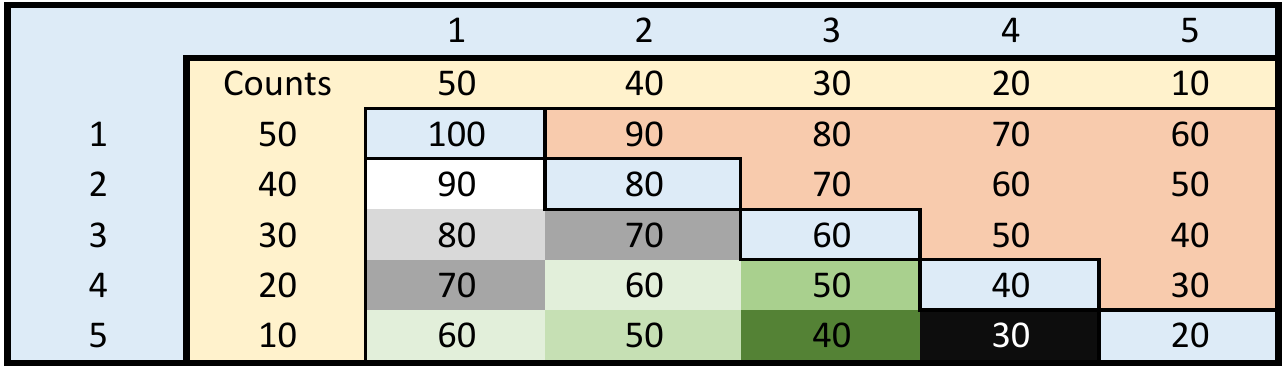}
\caption{In our example, the protein has 5 residues with mutations. The residues are sorted from largest to smallest (so residue 1 has the largest number of mutations, residue 2 the second largest number of mutations, etc.), and the inside of the table is calculated as the sum of the mutations on both residues. In the actual code, only the lower half of the table is considered and then only sequentially to decrease running time, but we present the whole table here for clarity.}
\label{mutationexample}
\end{center}
\end{figure}

To help illustrate this algorithm, consider a protein $N$ residues long with five mutated amino acid positions and suppose that we are interested in finding the two spheres that capture the most mutations. Without loss of generality and for ease of exposition, number the mutated residues 1 through 5. Further, assign the following mutation counts to these residues: residue 1 - 50 mutations, residue 2 - 40 mutations, residue 3 - 30 mutations, residue 4 - 20 mutations and residue 5 - 10 mutations. For clarity, we assume that the mutation counts are unique, but the algorithm is unaffected if there are identical mutation counts for some residues. We first construct the table shown in Figure \ref{mutationexample}, where the inside of the matrix is calculated to be the sum of the number of mutations at amino acid i and amino acid j. Observing, that the table is symmetric, we only need to evaluate the residues below the diagonal as the entries on the diagonal originate from residues that overlap each other perfectly. Thus for entry $(i,j)$ we are considering two spheres, one centered at residue $i$ and one centered at residue $j$. 

\begin{figure}[h!]	
\begin{center}
\includegraphics [width = 120mm] {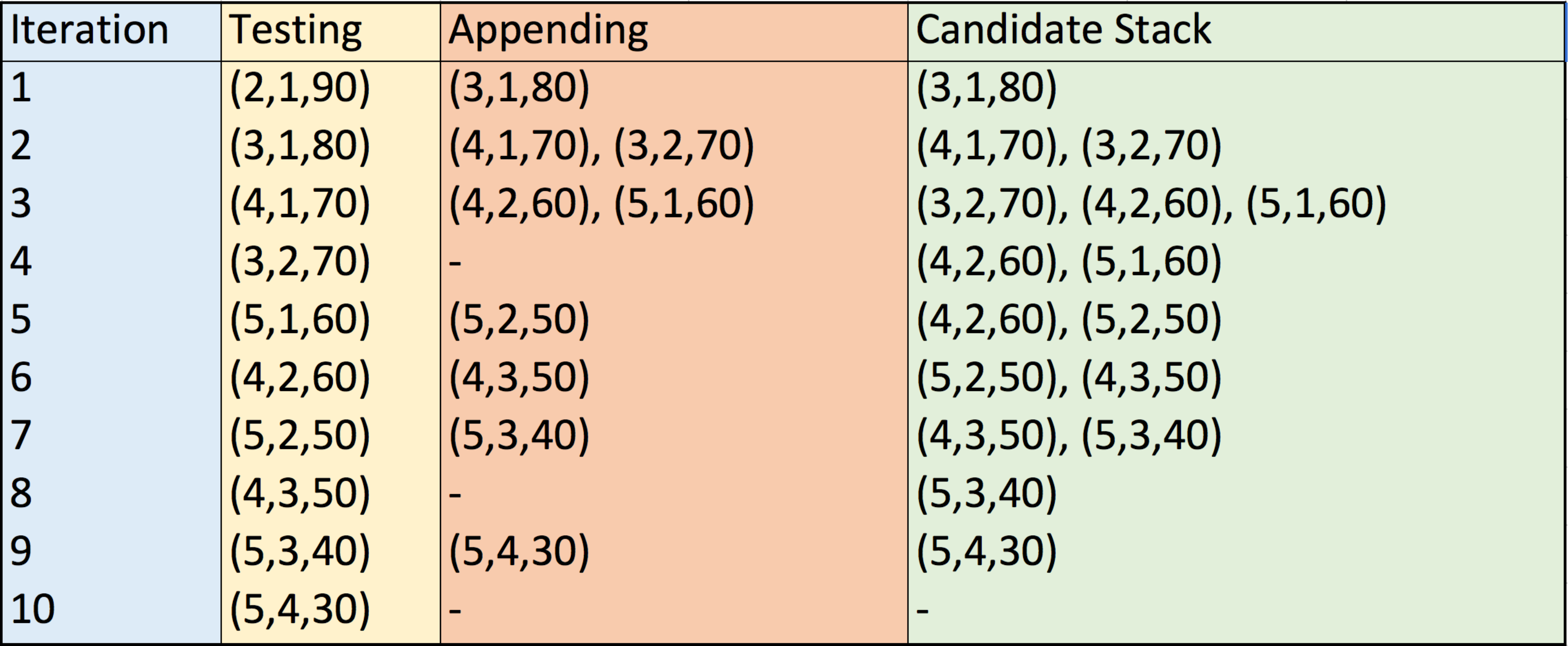}
\caption{This figure refers to the data in Figure \ref{mutationexample}. The first index, $i$, represents the row and the second index, $j$, represents the column. The third index, $s$, represents the total number of mutations at amino acids $i$ and $j$. Beginning in position (2,1,90), we then add (3,1,80) to the list, then \{(4,1,70), (3,2,70)\} and so forth. After each addition to the list, we pick the element with the highest value in the third position of the 3-tuple.}
\label{mutationexecution}
\end{center}
\end{figure}

The algorithm proceeds by appending to the ``candidate" stack the element directly below and the element directly to the right starting from the (2,1) entry. An ``element" consists of a 3-tuple $(i,j,s)$ where $i$ represents the row, $j$ represents the column and $s$ represents the value in position $(i,j)$.  After every two potential appends to the stack, the maximum value over the 3rd position is found (with the third position corresponding to the number of mutations covered by both spheres). The two spheres that contribute to this max element are then checked for overlap. If the spheres do \emph{not} overlap, then a successful case has been found and the algorithm completes. If the spheres \emph{do} overlap, the next  set of elements are appended and the process continues. By proceeding in the way described in Algorithm \ref{codesnip}, at each iteration, the pair of spheres being considered contain the maximum number of mutations possible from the remaining set of sphere combinations. Hence, once the first pair of non-overlapping spheres is found, the optimal sphere combination has been found and the algorithm can terminate. To see this process, see Figure \ref{mutationexecution}.

\subsection{Multiple Comparison Adjustment For Structures} \label{MultCompStruct}
A multiple comparison adjustment was performed to account for testing 1,110 protein structures. Since many structures may pertain to one protein, a Bonferroni adjustment is too conservative and an FDR approach was used. Specifically, a rough FDR (rFDR) \citep{gong_atlas_2009} approach, which is a good approximation to the standard FDR approach \citep{BH1995} when there are a large number of positively correlated or independent tests, was applied. In our case, the cutoff is set at:

$$rFDR = \alpha\left(\frac{k+1}{2k}\right)$$ where $k=1,110$, the total number of structures in the study. Using an $\alpha = 0.05$, the $rFDR \approx 0.025023$. To be conservative, we rounded down and deemed all clusters with a p-value less than or equal to $0.025$ to be significant. 

\section{Results and discussion}\label{discussion}

\begin{figure}[h!]
	\centering
	\includegraphics[scale=0.3]{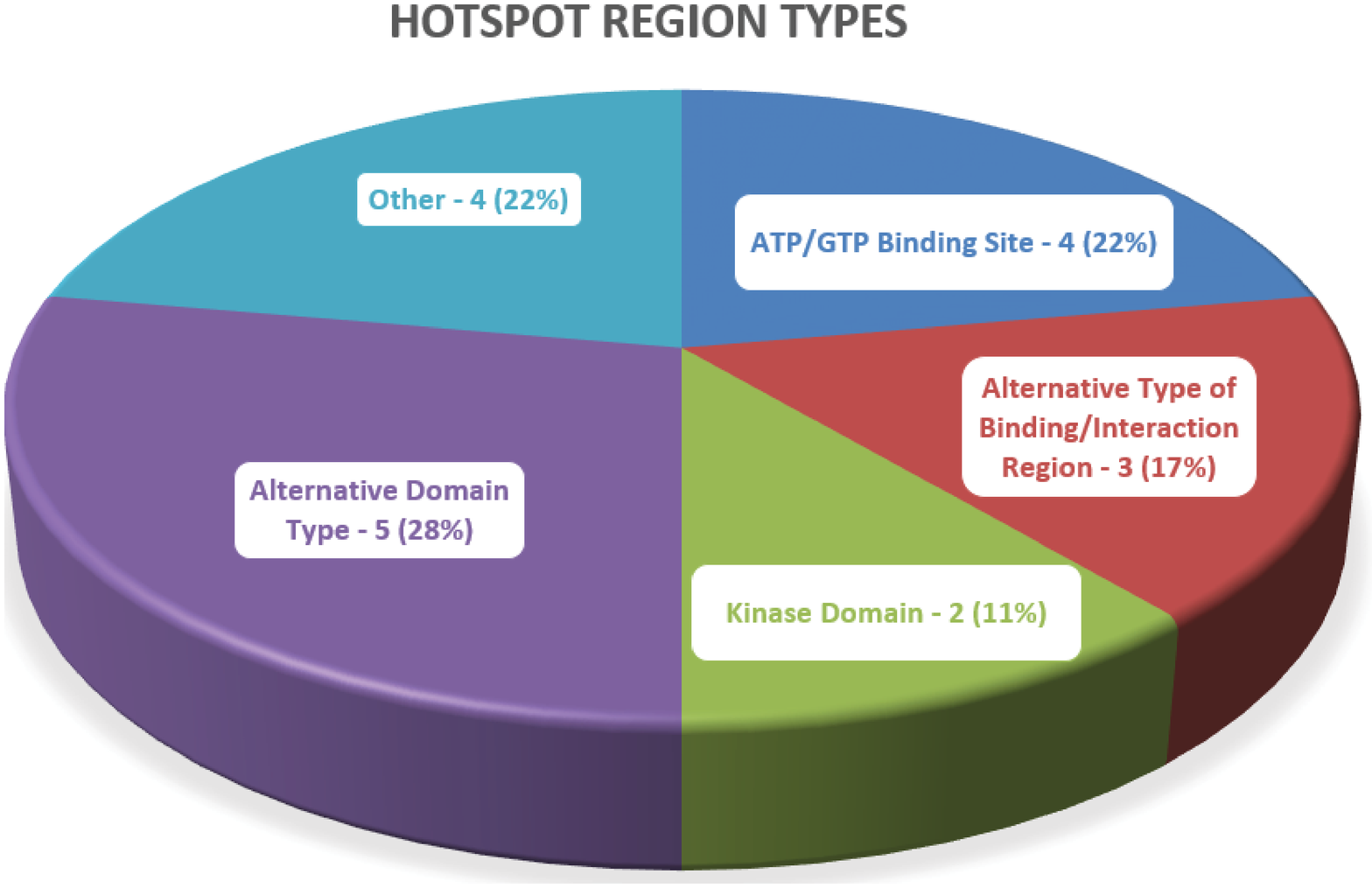}
	\caption{A breakout of what biologically relevant regions are overlapped by the most significant cluster for each of the 18 proteins. Overall, 77\% of the hotspots overlap a binding site or a protein domain.  For a full description regarding the overlap between \SpacePAC{} identified hotspots and structurally significant regions, see ``Relevant Sites" in the supplementary materials. }
	\label{fig:SiteBreakout}
\end{figure}

Of the 131 proteins considered, \SpacePAC{} identified 18 proteins with significant clustering as shown in Table \ref{tab:18SigTab}. For a full list of which structures were found significant under \SpacePAC{}, \GraphPAC{} and \NMC{},  see ``Results Summary" in the Supplementary materials. We note that while Table \ref{tab:18SigTab} shows the p-values for the 18 proteins found significant by \SpacePAC{}, there were 5 proteins that were found significant only by \GraphPAC{}, 1 protein only by \iPAC{} and 1 protein only by \NMC{} (see ``Results Table" in the supplementary materials for a complete list of what proteins were identified significant under each method). However, we also note that \SpacePAC{} identified the largest number of proteins with significant clustering at the same false positive threshold\footnote{The \GraphPAC{} algorithm was run using each of the three insertion methods described in \citep{GraphPAC_2013}. While the methodology is the same, the nature of the algorithm leads to different results when different insertion methods are considered. \SpacePAC{} outperformed \GraphPAC{} in comparison to each of the individual insertion methods.}. We further note that several of the proteins identified only by \SpacePAC{} have already been associated with cancer as shown in Section \ref{new}. Furthermore, as shown in Figure \ref{fig:SiteBreakout}, 14 out of the 18 proteins identified by \SpacePAC{} have their most significant hotspot overlap a biologically relevant region and three of the remaining four proteins (CTNNB1, FGFR3 and FSHR) have been implicated with cancer.

\begin{table}[h!tbp]
  \centering
 
    \begin{tabular}{|l|cccc|}
    \hline
          & \multicolumn{4}{c|}{\textbf{Method}}  \\
    \hline
    \textbf{Gene}  & \textbf{SpacePAC} & \textbf{iPAC}  & \textbf{GraphPAC} & \textbf{NMC} \\ \hline
    AKT1  & $<$1.00 E-03 & 4.48 E-04 & 5.54 E-04 & 5.55 E-04 \\
    ALK   & $<$1.00 E-03 & 1.99 E-42 & 3.89 E-35 & 2.16 E-21 \\
    BRAF  & $<$1.00 E-03 & $<$2.23 E-308 & $<$2.23 E-308 & $<$2.23 E-308 \\
    CHRM2 & 1.10 E-02 &       &       &  \\
    CTNNB1 & 2.00 E-03 & 6.69 E-03 & 1.09 E-03 & 1.09 E-03 \\
    DOCK2 & 2.20 E-02 & 9.29 E-03 & 3.39 E-03 &  \\
    FGFR3 & 2.00 E-02 &       &       &  \\
    FSHR  & 2.20 E-02 &       &       &  \\
    HRAS  & $<$1.00 E-3 & 1.55 E-23 & 1.87 E-32 & 2.68 E-15 \\
    IDE   & 2.00 E-02 & 3.60 E-03 &       &  \\
    IGF2R & 1.40 E-02 &       & 3.06 E-03 & 9.04 E-03 \\
    KIF18A & 6.00 E-03 & 1.56 E-02 & 1.56 E-02 & 1.56 E-02 \\
    KRAS  & $<$1.00 E-03 & $<$2.23 E-308 & $<$2.23 E-308 & $<$2.23 E-308 \\
    NRAS  & $<$1.00 E-03 & 1.53 E-75 & 6.65 E-77 & 6.77 E-77 \\
    PIK3CA & $<$1.00 E-03 & 4.73 E-118 & 4.73 E-118 & 4.73 E-118 \\
    PTEN  & $<$1.00 E-03 & 5.71 E-03 & 7.60 E-04 & 1.68 E-04 \\
    SEC23A & 1.70 E-02 &       & 1.18 E-02 &  \\
    TP53  & $<$1.00 E-3 & 1.78 E-134 & 6.22 E-169 & 1.08 E-88 \\
    \hline
    \end{tabular}%
  \caption{This table shows the p-value of the most significant cluster for each of the 18 proteins identified by \SpacePAC{} as well as the corresponding p-value under \iPAC{}, \GraphPAC{} and \NMC{}. A blank entry in position $(i,j)$ signifies that methodology $j$ did not find any structures with significant clustering for protein $i$. Note that  given $n_i$ total mutations for protein $i$, the pairwise methodologies perform $\frac{n_i(n_i-1)}{2}$ comparisons, one for each pair of mutations. As such, the p-values shown for \iPAC{}, \GraphPAC{} and \NMC{} have been multiplied by $\frac{n_i(n_i-1)}{2}$ in order to account for the multiple comparison and provide a number directly comparable to the \SpacePAC{} p-value. Also, while the \GraphPAC{} methodology was run under all three insertion methods (Cheapest, Nearest and Farthest) as described by \citep{GraphPAC_2013}, we display the minimum p-value over the three methods.}
  \label{tab:18SigTab}%
\end{table}

Specifically, for CTNNB1, the \SpacePAC{} identified hotspot covers mutations G34R and G34V, which are associated with hepatocellular carcinoma and hepatoblastoma \citep{legoix_beta-catenin_1999, koch_childhood_1999}, respectively.  Further, FSHR has been shown to be expressed in the vascular endothelial tissue of a wide range of human tumors including lung, breast, prostate, colon and leiomyosarcoma \citep{siraj_expression_2013}. For more detail on FGFR3, see Section \ref{new}.

Finally, we evaluated \SpacePAC{} performance via two common machine learning methods, \emph{PolyPhen-2} \citep{adzhubei_method_2010} and \emph{CHASM} \citep{carter_cancer-specific_2009}. Before we summarize the results, we note that \emph{PolyPhen-2} and \emph{CHASM} utilize a large set of features when evaluating each mutation. The advantage of \SpacePAC{} is that it is able to be run with vastly less a priori information. 
Out of the 38 mutated amino acids that fall within \SpacePAC{} identified hotspots, PolyPhen-2 identifies 36 (95\%) as damaging while CHASM identifies 31 (82\%) as driver mutations at a FDR of 20\%. On the protein level, PolyPhen-2 identifies all the 18 proteins identified by \SpacePAC{} as significant while CHASM identifies 14 proteins as significant. Moreover, \SpacePAC{} identifies several proteins with significant clustering that are missed by the machine learning methods. For instance, \SpacePAC{} identifies FSHR as significant, which, as described above, has recently been associated with cancer. However, \CHASM{} calculates a FDR of 0.45 for FSHR, which is above the significance threshold. See ``Performance Evaluation" in the supplementary materials for more information. 

\subsection{\SpacePAC{} finds novel proteins} \label{new}
As described in Section \ref{discussion}, \SpacePAC{} identified three novel proteins that were missed by \NMC{}, \iPAC{} and \GraphPAC{}. We will now consider two of these proteins, Fibroblast Growth Factor Receptor 3 (FGFR3), and  Muscarinic Acetylcholine Receptor M2 (CHRM2 or M2).

\begin{figure}[h!]
	\centering
	\includegraphics[scale=0.08]{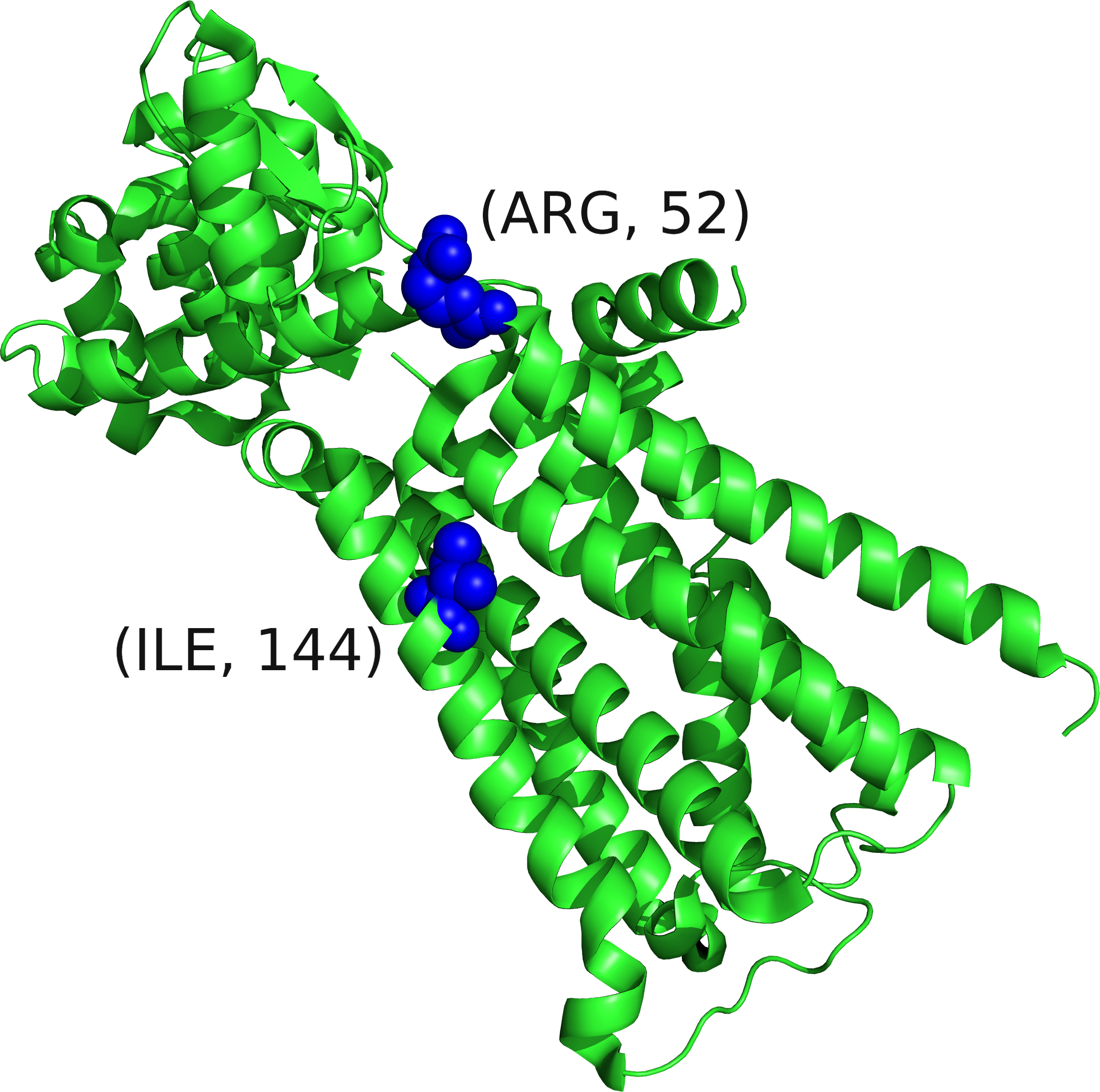}
	\caption{The CHRM2 structure (PDB ID: 3UON) where residues 52 and 144 are highlighted.}
	\label{fig:CHRM2Cluster}
\end{figure}

The CHRM2 structure (PDB ID: 3UON) \citep{haga_structure_2012} was identified by \SpacePAC{} as having two significant mutational hotspots (p-value = $0.011$) located at residues 52 and 144 (see Figure \ref{fig:CHRM2Cluster}). CHRM2, essential for the physiological regulation of cardiovascular function \citep{haga_structure_2012} has been implicated in a variety of cardiovascular diseases. Recently, CHRM2 has also been associated with both autoimmune diseases and cancer \citep{ockenga_non-neuronal_2013}. Current research shows that M2 receptors are expressed in both glioblastoma cell lines and human samples. Moreover, the M2 agonist arecaidine strongly decreases cell proliferation in both primary cultures and cell lines in a dose and time dependent response profile. This suggests that M2 activation has an important role in suppressing glioma cell proliferation and can provide a novel therapeutic target \citep{ferretti_m2_2013}. Had the spatial structure not been taken into account, as under \NMC{},  or if the structure was accounted for but only via remapping to 1D space, as under \iPAC{} and \GraphPAC{}, this cluster would have been missed.

\begin{figure}[h!]
	\centering
	\includegraphics[scale=0.08]{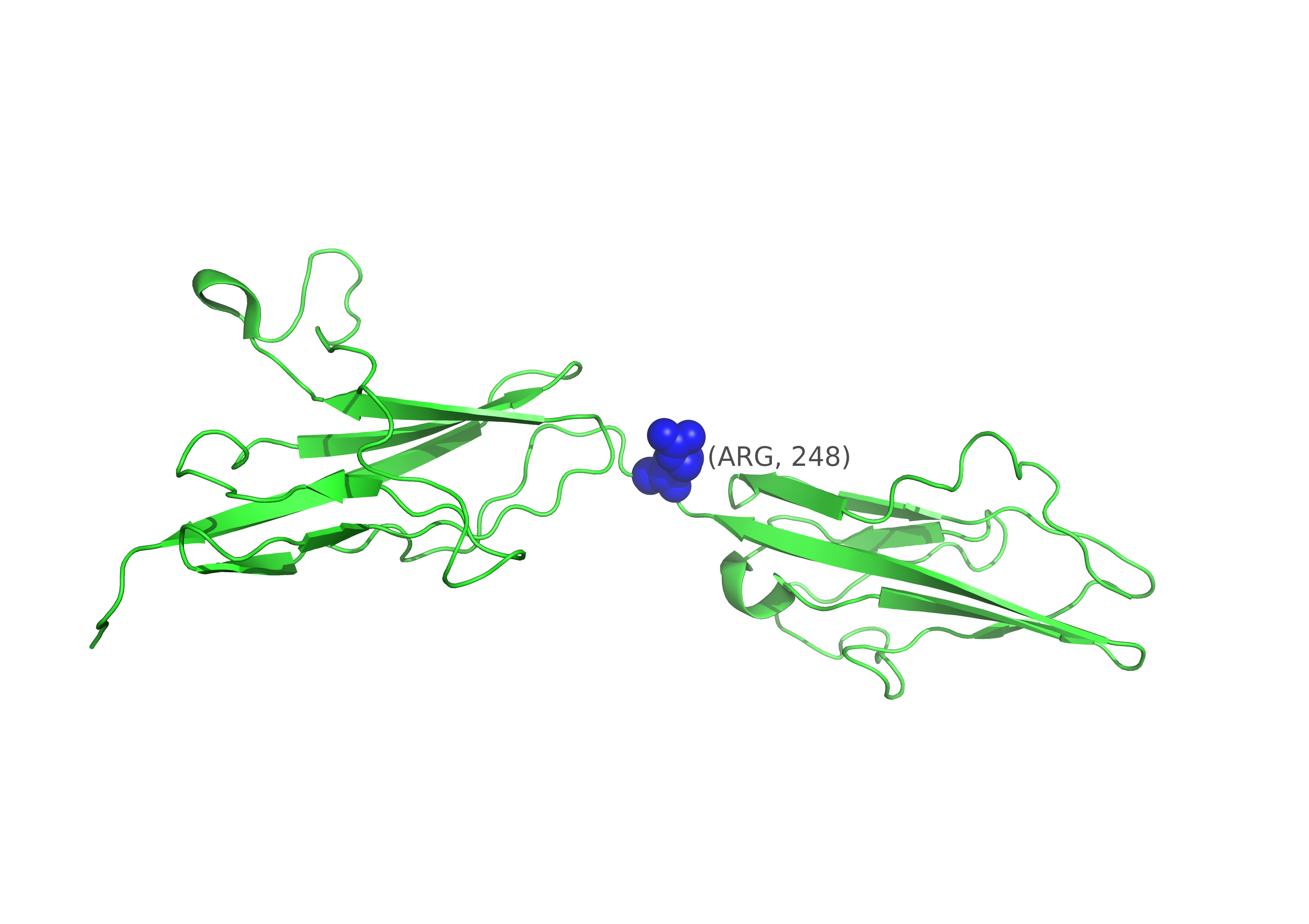}
	\caption{The FGFR3 structure (PDB ID: 1RY7) where residue 248 is highlighted blue. }
	\label{fig:FGFR3Cluster}
\end{figure}

\SpacePAC{} identified the FGFR3 structure (PDB ID: 1RY7) \citep{olsen_insights_2004} as having one significant hotspot (p-value $= 0.020$) centered at amino acid 248 (see Figure \ref{fig:FGFR3Cluster}). FGFR3 is a tyrosine-protein kinase which plays a critical role in regulating cell differentiation, proliferation and apoptosis and is often associated with cancer and developmental disorders \citep{hart_identification_2001}. Mutation R248C occurs in the Ig-like domain 
 and is a severe and lethal mutation associated with bladder cancer \citep{hadari_fgfr3-targeted_2009} along with a variety of other phenotypes such as thanatophoric dwarfism \citep{rousseau_missense_1996} and epidermal nevi \citep{hafner_mosaicism_2006}. This cluster represents a perfect example of signal loss when all pairwise mutations are considered. In the case of our data, as all the mutations occur on one residue, a cluster is formed at that one residue only. There is therefore no difference between any of the pairwise methods as the remapping step has no effect. However, since \iPAC{}, \GraphPAC{} and \NMC{} need to account for all pairwise comparisons between mutations (all occurring on residue 248), the signal is lost under all three methods. On the other hand, as \SpacePAC{} does not need to perform such a correction, it is successfully able to detect the cluster. Moreover, we note that \citet{qing_antibody-based_2009} recently developed an anti-FGFR3 monoclonal antibody that interferes with FGFR3 binding and inhibits R248C \citep{hadari_fgfr3-targeted_2009}.

\subsection{\SpacePAC{} improves cluster localization} \label{localization}
For protein-structure combinations in which mutational clusters are detected by other methods, \SpacePAC{} is often able to provide a smaller set of clusters compared to the pairwise methods while still covering the majority of mutations. To illustrate this point we consider two examples, the BRAF structure (PDB ID: 3Q96)  \citep{gould_design_2011} where \SpacePAC{} identifies 3 mutational hotspots and the ALK structure (PDB ID: 2XBA) \citep{bossi_crystal_2010} where \SpacePAC{} identifies 2 mutational hotspots.

BRAF is a well known oncogene that is part of the RAS-RAF-MEK-ERK-MAP kinase pathway which is often activated in human tumors. Further, it is estimated that approximately 90\% of mutations in this gene are a substitution of a glutamate for a valine at residue 600 (V600E) \citep{tan_detection_2008}. In our mtuational data, 187 (83.5\%) mutations were on residue 600 with the remaining 37 mutations spread over 13 other residues.  Mutations on V600 typically result in constitutively elevated kinase activity and have been found in a wide range of cancers such as metastatic melanoma, ovarian carcinoma and colorectal carcinoma \citep{davies_mutations_2002, rajagopalan_2002, hingorani_2003, sjoblom_consensus_2006, greenman_patterns_2007, andreu-perez_protein_2011}. Due to the large number of V600 mutations, \SpacePAC{} and all the pairwise methods identified residue 600 as the most significant ``cluster" in all structures where tertiary information was available on that residue. It is worth noting that BRAF V600 inhibitors, such as Vemurafenib, have already been developed, further supporting the hypothesis that mutational clusters may represent pharmaceutical targets \citep{mao_resistance_2012}.

As the signal presented by V600 is so strong, it may mask the signal from other mutations within the BRAF protein. As such, we considered structure 3Q96 which does not have tertiary information for residues 600 and 601. Of the remaining 28 mutations spread over 12 residues, \SpacePAC{} identifies three hot spots with 7-8 mutations per cluster as shown in Table \ref{BRAFSig} (combined hot spot p-value $<$1.00 E-03). Moreover, each of the three regions identified has been associated with oncogenic elevated kinase activity as well as a variety of cancers such as lung adenocarcinoma, melanoma, colorectal adenocarcinoma and ovarian serous carcinoma \citep{ davies_mutations_2002, lee_braf_2003, greenman_patterns_2007}.

\begin{table} [h!]
	\begin{center}
    \begin{tabular}{|c|c|c|c|}
    \hline
    Hotspot & Center & Within Sphere & \# Mutations \\ 
    \hline
    A       & 465    & (464, 465, 466) & 8           \\
    B       & 470    & (469, 470, 471) & 7          \\
    C       & 596    & (595, 596, 597) & 7          \\
    \hline
    \end{tabular}
    \end{center}
    \caption{The three hot spots identified by \SpacePAC{} for the BRAF structure (PDB ID: 3Q96) at an optimal radius of 4\AA. See Figure \ref{fig:BRAFCluster} for a visual orientation.}
    \label{BRAFSig}
\end{table}

Together, the three hotspots identified by \SpacePAC{} cover 79\% of the mutations for which tertiary information is available. Moreoever, while \NMC{}, \iPAC{} and the three \GraphPAC{} methods report approximately 8 to 16 times as many clusters as \SpacePAC{} (see Table \ref{BRAFCovered}), the additional clusters only cover the remaining 21\% of mutations. Finally, all the residues that do not fall within \SpacePAC{} hot spots are those which have only one mutation.  These additional clusters stem from the fact that \NMC{}, \iPAC{} and \GraphPAC{} must consider every pairwise combination of mutations resulting in many clusters that only differ from each other by a few residues. Further, by considering every pairwise combination, many smaller clusters are often combined into larger clusters with a less significant p-value. While technically still a ``significant" cluster, these extra clusters provide little additional information. As \SpacePAC{} does not have to consider every combination, it does not suffer from this issue.

 \begin{figure}[h!]
	\centering
	\includegraphics[scale=0.08]{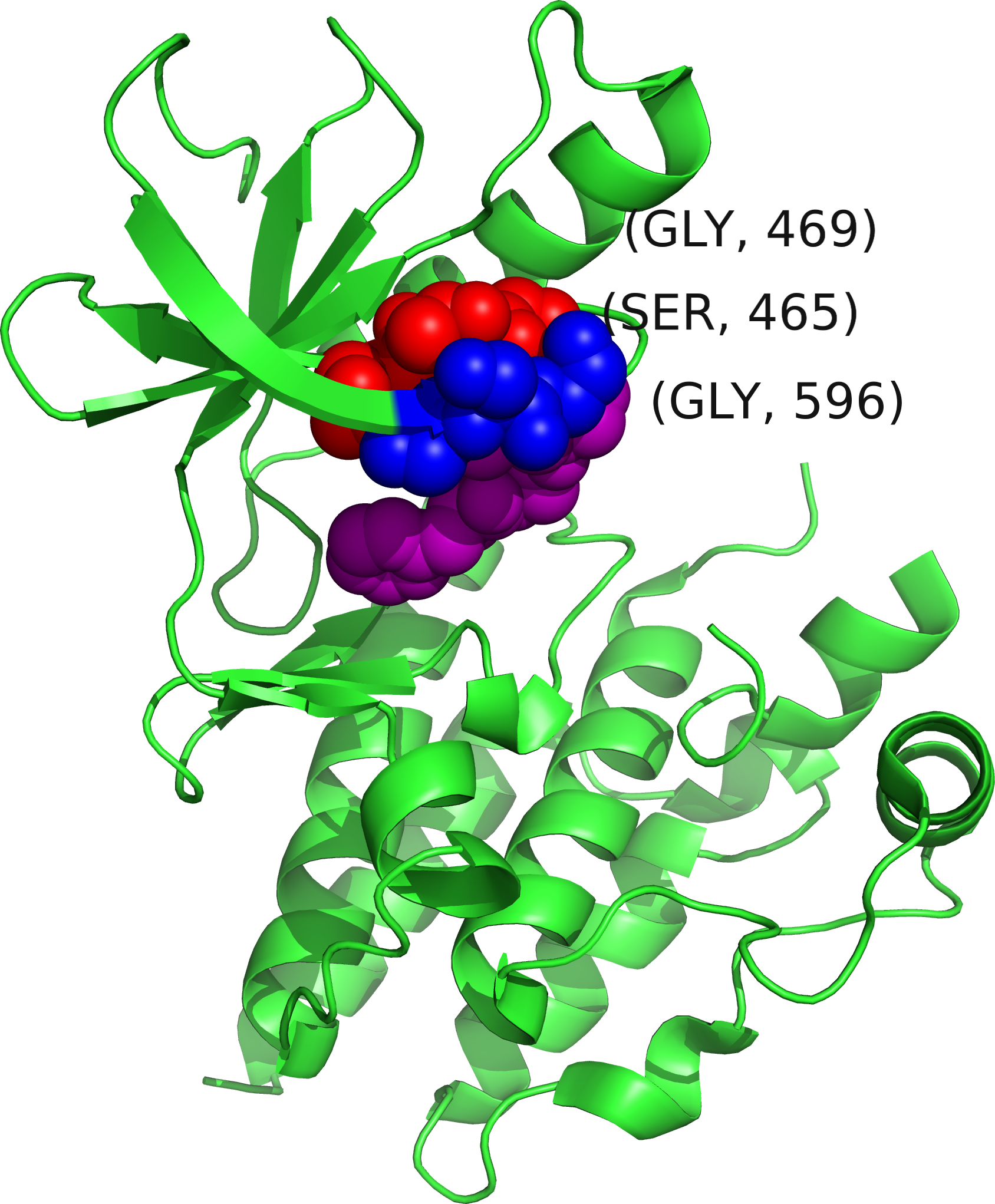}
	\caption{The BRAF structure (PDB ID: 3Q96) where cluster 464-466 is shown in blue, 469-471 is shown in red and 595-597 is shown in purple. The central residue in each cluster (465, 470 and 596 for the blue, red and purple clusters respectively) is labeled. }
	\label{fig:BRAFCluster}
\end{figure}

We now consider the Anaplastic Lymphoma Kinase (ALK) protein for which \SpacePAC{} identifies two mutational hotspots (combined hot spot p-value $<$0.001) (see Table \ref{ALKSig}). The ALK protein is a receptor-type tyrosine kinase that is preferentially expressed in neurons during the late embryonic stages \citep{motegi_alk_2004}. Mutations in this protein have been associated with both neuroblastoma as well as non-small cell lung cancer \citep{bang_treatment_2012, george_activating_2008}. Hotspots A and B in Table \ref{ALKSig} both occur in the protein kinase. Further, mutations on F1174 and R1275 can cause constitutive activation which impairs receptor trafficking \citep{mazot_constitutive_2011}. We note that \SpacePAC{} perfectly identifies both hotspot locations. Moreover, it has recently been shown that activating mutations F1174L and R1275Q provide therapeutic targets in neuroblastoma \citep{george_activating_2008}, supporting the hypothesis that mutational clusters are indicative of activating mutations.

\begin{table} [h!]
	\begin{center}
    \begin{tabular}{|c|c|c|c|}
    \hline
    Hotspot & Center & Within Sphere & \# Mutations \\ 
    \hline
    A       & 1174    & (1173, 1174, 1175) & 11           \\
    B       & 1275    & (1274, 1275, 1276) & 12          \\
    \hline
    \end{tabular}
    \end{center}
    \caption{The two hot spots identified by \SpacePAC{} as significant for the ALK Structure (PDB ID: 2XBA) \citep{bossi_crystal_2010} at an optimal radius of 4\AA. See Figure \ref{fig:ALKCluster} for a visual orientation.}
    \label{ALKSig}
\end{table} 

The two hotspots identified by \SpacePAC{} cover 88.5\% of all the mutations in our data with the remaining mutations occurring on residues with only one mutation each. Further, as seen in Table \ref{BRAFCovered}, the pairwise methods have 3 to 5.5 times as many clusters as \SpacePAC{}. As before, by not having to consider every pairwise combination and thus not reporting similar clusters, \SpacePAC{} is able to better localize the critical mutational areas.

 \begin{figure}[h!]
	\centering
	\includegraphics[scale=0.08]{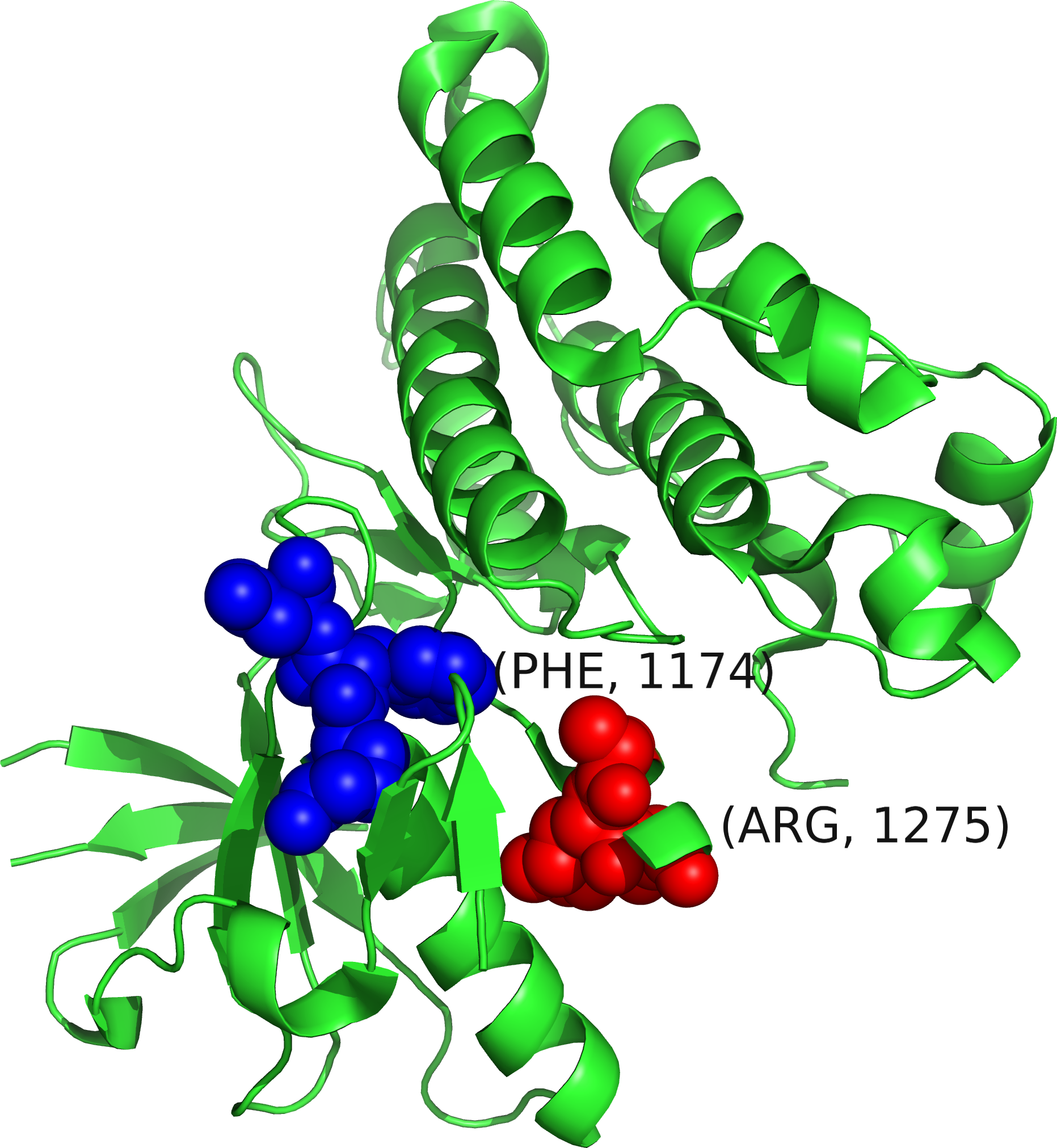}
	\caption{The ALK structure (PDB ID: 2XBA) where cluster 1173-1175 is shown in blue and cluster 1274-1276 is shown in red. The central residue in each cluster (1174 and 1275 for the blue and red clusters respectively) is labeled. }
	\label{fig:ALKCluster}
\end{figure}

\begin{table}[h!]
	\begin{center}
	\begin{tabular}{|c|c|c|}
	\hline
	& \multicolumn{2}{c|}{\# Clusters/Hotspots}\\
	\hline
	Method/ Protein-Structure &  BRAF - 3Q96 & ALK-2XBA \\
	\hline
	\SpacePAC{}	& 3 & 2 \\
	\NMC{} & 22 &  11 \\
	\iPAC{} & 45 & 11\\
	\GraphPAC{}-Cheapest & 36 & 7\\
	\GraphPAC{}-Nearest & 45 & 8 \\
	\GraphPAC{}-Farthest & 47 & 6 \\
	\hline
	\end{tabular}
	\end{center}
	\caption{The number of clusters found under each method. As \SpacePAC{} does not need to consider every pairwise combination of mutations, the method provides a much smaller number of potential hotspots while still covering the majority of mutations.}	
	\label{BRAFCovered}
\end{table}

\section{Conclusion}

In this article we provide a novel algorithm to account for protein tertiary structure when identifying mutational clusters in proteins. By considering the protein structure directly in 3D space, we avoid the use of a dimension reduction algorithm and potential information loss. Further, by not considering every pair of mutations, we are able to reduce the multiple comparison penalty and identify additional clusters. We show several examples of clusters that are not identified by alternative methods as well as the ability to improve cluster localization while still covering the majority of mutations.  Moreover, several of our examples identified clusters which overlap potential therapeutic targets, supporting the hypothesis that clusters may be indicative of activating driver mutations. Finally, since the \SpacePAC{} methodology does not need to look at every pairwise combination of mutations it also runs much faster for proteins with many ($>400$) mutations. In these situations, while the pairwise methods may take several days to complete, \SpacePAC{} still finishes in a matter of hours. For proteins with fewer mutations, the running time of all the pairwise methods as well as \SpacePAC{} is comparable, with the majority of protein/structure combinations terminating in under 10 minutes when executed on a consumer desktop with an Intel i7-2600k processor (at a frequency of 3.40 GHZ) and 16 GB of DDR3 RAM.

\SpacePAC{}, while presenting an important alternative to the one dimensionality restriction required by \iPAC{} and \GraphPAC{}, is nonetheless subject to several limitations. First,  \SpacePAC{} is currently limited to at most three mutational hotspots to save on computational time. While it is unlikely that a single structure will have more than three hotspots, the extension to allow \SpacePAC{} to account for more than three hotspots is algorithmically simple. As \SpacePAC{} was able to process our entire database of protein/structure combinations in under 5 hours (with all the structures evaluated in parallel), this restriction is minimal and will only grow smaller as computational power increases.

Second, to satisfy the uniformity assumption, the mutational status of each amino acid must be known. However, due to improvements in high-throughput sequencing, this is rapidly becoming a non-issue.  Next, unequal rates of mutagenesis in specific genomic regions may violate the assumption that each residue has an equal probability of mutation. To help ensure that our data met this statistical assumption, we only considered missense mutations as many insertion and deletion mutations are sequence dependent.  Relatedly, while the literature shows that CpG dinucleotides often have a mutational rate ten times or higher when compared to other locations \citep{Sved01061990},  approximately only 14\% of the clusters presented in section \ref{new} and \ref{localization} overlapped CpG sites. Similarly, cigarette use typically causes transversion mutations within lung carcinomas \citep{ye_2010} while colorectal carcinomas result in transition mutations \citep{hollstein_p53_1991}. In the case of KRAS however, the vast majority of mutations are located on residues 12, 13 and 61 for both  cancers. This signifies that while the mutational type may differ, the impact on mutation location is minimal and does not violate the uniformity assumption.  Lastly, it is worth noting that as we obtained our mutational data from the COSMIC database, specific tissue types may be more represented than others. However, under this situation our analysis would be more conservative and the resulting findings even more significant as explained below. 
Aggregating over all tissues increases the total number of mutations, thus increasing the number of simulated mutations within any given segment of the protein. The resulting p-value of any observed hotspot would lose significance as more mutations would be simulated within the cluster. Overall, while this as well as previous studies are impacted by several external factors, it appears that selection of the cancer phenotype is the primary cause of clustering.

In conclusion, \SpacePAC{} presents a novel approach to account for protein tertiary structure when identifying mutational hotspots. We show that \SpacePAC{} identifies novel clusters of biological relevance, improves cluster localization and in several cases identifies pharmaceutical targets for which therapies are already in production. In turn, we further confirm the hypothesis that mutational hotspots are indicative of driver mutations and show that \SpacePAC{} can be used to quickly locate such potential mutations as additional structures are published.

\bigskip

\section*{Author's contributions}
 GR, DZ and HZ developed the \SpacePAC{} methology. KC and YC were responsible for obtaining the mutation data from the COSMIC database. GR and YC executed the methodology on the protein structures. GR drafted the original manuscript while KC, YC, YM, DZ and HZ were responsible for revisions. HZ finalized the manuscript. All authors have read and approved the final text.   This work was supported in part by NSF Grant DMS 1106738 (GR, HZ), NIH Grants GM59507 and CA154295 (HZ).

\section*{Acknowledgements}
We thank Dr. Francesca Chiaromonte for her time and help in discussing this methodology.
  
\section*{Competing Interests}
   The authors declare that they have no competing interests.

 \bibliographystyle{natbib}  
 \bibliography{article}      


\end{document}